\documentclass[12pt,preprint]{aastex}
\usepackage{epsfig}
\usepackage{natbib}
\usepackage{graphicx}
\usepackage{slashbox}
\usepackage{multirow}
\usepackage{lscape}
\usepackage{mathrsfs,amssymb}
\usepackage{amsmath}
\newcommand	\beq	{\begin{equation}}	%{\begin{displaymath}}
\newcommand	\eeq	{\end{equation}}	%{\end{displaymath}}
\newcommand       \Angstrom     {\,{\rm \AA}}
\newcommand       \AU           {\,{\rm AU}}
\newcommand       \cm           {\,{\rm cm}}

\newcommand       \erg          {\,{\rm erg}}

\newcommand       \K            {\,{\rm K}}
\newcommand       \pc           {\,{\rm pc}}

\newcommand       \s            {\,{\rm s}}

\newcommand       \yr       {\,{\rm yr}}

\newcommand       \Myr      {\,{\rm Myr}}

\newcommand       \simlt        {\lesssim}

\newcommand       \gtsim        {\gtrsim}
\newcommand       \ltsim        {\lesssim}

\newcommand       \mum          {\,{\rm \mu m}}

\newcommand       \Teff         {T_{\rm eff}}

\newcommand       \Lsun         {\,{L_\odot}}

\newcommand       \Lstar        {L_\star}

\newcommand       \simali       {\sim\,}
\newcommand       \magni        {\,{\rm mag}}

\def    \amin		{a_{\rm min}}
\def    \amax		{a_{\rm max}}

\newcommand{\Msun}      {\,{M_\odot}}
\newcommand{\Mearth}    {\,{M_\earth}}

\newcommand       \apahmin      {a^{\rm PAH}_{\rm min}}
\newcommand	  \rmin	        {r_{\rm min}}
\newcommand	  \rmax	        {r_{\rm max}}
\newcommand	  \rp           {r_{\rm p}}
\shorttitle{Dust and PAHs in HD\,34700}
\shortauthors{Seok \& Li}
\title{
Dust and Polycyclic Aromatic Hydrocarbon 
in the HD\,34700\ Debris Disk
}
\author{Ji Yeon Seok and Aigen Li}
\affil{Department of Physics and Astronomy,
        University of Missouri,
        Columbia, MO 65211, USA;
        {\sf seokji@missouri.edu,
             lia@missouri.edu}}

\begin{document}
%\defcitealias{Li2003}{LL03}

\begin{abstract}
The debris disk around 
the Vega-type star HD\,34700
is detected in dust thermal emission 
from the near infrared (IR) to millimeter (mm)
and submm wavelength range. 
Also detected is a distinct set
of emission features at 3.3, 6.2, 
7.7, 8.6, 11.3 and 12.7$\mum$, 
which are commonly attributed
to polycyclic aromatic hydrocarbon 
(PAH) molecules.
We model the observed dust IR 
spectral energy distribution (SED)
and PAH emission features  
of the HD\,34700 disk in terms of
porous dust and astronomical-PAHs.
Porous dust together with 
a mixture of neutral and ionized PAHs 
closely explains the dust IR SED 
and PAH emission features 
observed in the HD\,34700 disk.
Due to the stellar radiation pressure
and Poynting--Robertson drag together with
the photodissociation of PAHs, 
substantial removal of dust and PAHs 
has occurred in the disk, and
continuous replenishment of these materials is 
required to maintain their current abundances.
This implies that these materials are
not primitive but secondary products probably
originating from mutual collisions 
among planetesimals, asteroids, and comets.

\end{abstract}


\section{Introduction}\label{sec:intro}
Vega-type stars are main-sequence (MS) stars 
with infrared (IR) emission in excess of
the stellar photospheric radiation 
(see Backman \& Paresce 1993).
This ``excess'' emission is attributed to
the thermal emission of solid dust grains
in a ``debris'' disk,
an optically thin disk orbiting around
the central star. 
In debris disks, the gas is tenuous,
and dust grains are removed easily 
by radiation pressure (RP) and the Poynting--Roberson drag
(Burns et al. 1979, Backman \& Paresce 1993).
To sustain a debris disk, dust grains need 
to be replenished continuously by collisions 
between planetesimals, asteroids, and comets.
This implies that grains in a debris disk
are not the primordial materials left over from
the protostellar molecular cloud 
but remnants of the planet formation process.

HD\,34700, also known as SAO\,112630, is classified 
as a Vega-like star based on its IR excess 
(e.g., see Walker \& Wolstencroft 1988). 
It has a spectral type of G0\,IVe, 
an effective temperature of $T_{\rm eff}\approx6000\K$
(see Sch\"utz et al.\ 2009 and references therein),
and a total stellar luminosity of
$\Lstar\approx20.4\,\Lsun$ 
(Acke \& van den Ancker 2004).
With a Hipparcos parallax of $\approx0.86\pm1.84$\,mas,
its distance $d$ is very uncertain,
which makes the estimate of its age, 
mass, and size ambiguous.
It has been suggested that $d$ is 
at least $\simali$100$\pc$ (Torres 2004),
and its apparent proximity to Orion provides
an upper limit of $d\sim430\pc$. 
Assuming $d\approx200\pc$, Torres (2004)
estimated the stellar mass to be
$M_\star$\,$\approx$\,1.1--1.2 $\Msun$.

The HD\,34700 system is known as 
a multiple system consisting of 
a spectroscopic binary (Torres 2004)
and two other faint stellar components 
which are at a separation of 
$\simali$5\farcs2 and $\simali$9\farcs2,
respectively, for Component B and  Component C
(Sterzik et al.\ 2005).
While the two faint components are classified 
as bonafide T Tauri stars (TTS) based on their 
strong H$\alpha$ emission and \ion{Li}{1} 
$\lambda6798$ absorption,
the nature of the spectroscopic binary 
is somewhat vague.
Both the H$\alpha$ emission and
the \ion{Li}{1} $\lambda6798$ absorption
(with an equivalent width of $\simali$0.6 
and $\simali$0.17$\Angstrom$, respectively) 
detected in HD\,34700 are relatively weak
(Arellano Ferro \& Giridhar 2003, Sterzik et al.\ 2005),
indicating that the age of HD\,34700 
cannot be a few mega-years 
but a few tens of mega-years (Torres 2004).
Moreover, as we will show later in 
Section~\ref{sec:discussion} that the lifetime of 
the dust grains in the disk around HD\,34700
is shorter than the stellar age,
therefore it is plausible that HD\,34700 
has a debris disk rather than 
a protoplanetary disk.

The HD\,34700 debris disk has been 
extensively studied observationally, 
including space-borne and ground-based 
broadband photometry of
the IR ``excess'' emission
from the near-IR to submillimeter wavelengths.
Of particular interest is the detection of
the prominent 3.3, 6.2, 7.7, 8.6, and 11.3$\mum$ 
emission bands in HD\,34700.
These emission features are collectively 
known as the ``unidentified infrared'' (UIR)
emission bands (see Tielens 2008, 
Peeters 2014, Joblin 2015)
and are now generally attributed to
polycyclic aromatic hydrocarbon (PAH) molecules 
(L\'{e}ger \& Puget 1984; 
Allamandola et al. 1985).  
They are ubiquitously seen in a wide
variety of Galactic and extragalactic regions,
including protoplanetary disks 
around Herbig Ae/Be stars and TTS
and disks in transition from a primordial to
debris phase 
(e.g., see Brooke et al.\ 1993, 
Siebenmorgen et al.\ 2000,
Acke \& van den Ancker 2004,
van Boekel et al.\ 2004, 
Geers et al.\ 2005, 
Sloan et al.\ 2005,
Weinberger \& Becklin 2005, 
Furlan et al.\ 2006, 
Habart et al.\ 2006, 
Keller et al.\ 2008).
However, the detection of the PAH emission features 
in debris disks is relatively rare. 
To the best of our knowledge,
they are reportedly seen in only a few
debris disks (Coulson \& Walther 1995, 
Sylvester et al. 1997,
Smith et al. 2004).
For example, Acke et al.\ (2010) found that
37 out of 53 Herbig Ae/Be stars show PAH features 
in their IR spectra obtained with
the {\it Infrared Spectrograph} (IRS)
on board the \textit{Spitzer Space Telescope}.
In contrast, Chen et al.\ (2006) reported no detection 
of PAH emission in the {\it Spitzer}/IRS spectra of 
59 MS stars classified as Vega-like objects 
based on the \textit{IRAS} 60$\mum$ excess.
%

%{\bf Despite of the lack of detection in debris disks,}
PAHs play an important role 
in gas-rich planet-forming disks.
They dominate the heating of the gas
in their surface layers.
They are also diagnostic of the presence of 
very small grains in their surface layers,
which in return are diagnostic of the settling 
and coagulation processes occurring in the disks 
that ultimately lead to the formation of planetesimals 
(see J.\,Y.~Seok \& A.~Li 2015, in preparation). 
For gas-depleted debris disks, PAHs could potentially
be used to probe their disk structures
that are widely considered as signposts 
for planet formation.
Due to their stochastic heating nature (see Draine \& Li 2001), 
the PAH emission features are expected to be spatially more 
extended than the emission of large grains 
that are in thermal equilibrium with the stellar radiation. 
Therefore, debris disks could be
more easily resolved at the PAH emission bands.

The 3.3$\mum$ PAH C--H stretching feature was clearly 
detected in the $\simali$1.9--4.2$\mum$ spectrum 
of HD\,34700 obtained with the medium-resolution spectrograph
{\it SpeX} at the NASA {\it Infrared Telescope Facility} 
(IRTF; Smith et al.\ 2004).
The 6.2 and 7.7$\mum$ C--C stretching features
as well as the 11.3$\mum$ out-of-plane
C--H bending feature were detected
in the $\simali$2.5--12$\mum$ spectrum 
of HD\,34700 obtained by the spectrophotometer
{\it ISOPHOT} on board 
the {\it Infrared Space Observatory} (ISO)
albeit its limited spectral resolution 
of $\lambda/\Delta\lambda\approx90$
(Acke \& van den Ancker 2004),
and later by \textit{Spitzer}/IRS 
with higher sensitivities 
(Bern\'e et al.\ 2009, Sch\"utz et al.\ 2009).
By combining the \textit{ESO}/TIMMIS2 
$\simali$8--13$\mum$ N-band spectrum with 
the \textit{Spitzer}/IRS spectrum, 
Sch\"utz et al.\ (2009) also reported
marginal detections of the 8.6$\mum$ 
C--H in-plane bending feature
and the 12.7$\mum$ out-of-plane
C--H bending feature.

With a distinct set of prominent PAH emission 
features detected, HD\,34700 offers a rare 
opportunity to explore the nature and origin 
of PAHs in debris disks. In this work we model
the dust and PAH emission of the HD\,34700 debris disk,
with an aim of exploring its dust and PAH properties.
This paper is organized as follows.
We describe the observational data
of dust and PAH emission in Section \ref{sec:data}.  
We present the PAH and dust models 
in Section \ref{sec:model}. 
The results are summarized in 
Section \ref{sec:results} and discussed 
in Section \ref{sec:discussion} with special
attention paid to the photophysics of PAHs. 
The major conclusions are summarized 
in Section \ref{sec:summary}.

%%%%%%%% Table 1: SED %%%%%%%%
\begin{deluxetable}{ccccc}
\center
\tablecaption{\footnotesize
              \label{tab:flx}
              Photometric Data 
              for the HD\,34700 System
              }
\tablehead{\colhead{Telescope} &\colhead{Wavelength} 
& \colhead{Flux Density} 
& \colhead{Uncertainty}  \\ 
  \colhead{/Mission} & \colhead{($\mu$m)} 
& \colhead{(Jy)} & \colhead{(Jy)} } 
\startdata
 & 0.36 ($U$)& 0.239 & 0.0088  \\ 
 & 0.44 ($B$) & 0.578 & 0.0267 \\
 EXPORT & 0.53 ($V$) & 0.887 & 0.0409  \\
 & 0.69 ($R$)& 0.941 & 0.0780  \\
 & 0.83 ($I$) & 0.887  & 0.0818  \\
%2MASS & JHK \\
\tableline
 & 1.235 ($J$) & 0.968 & 0.0205  \\
 2MASS & 1.662 ($H$) & 0.847 & 0.0179 \\
 & 2.159 ($K_S$) & 0.678 & 0.0150  \\
%WISE \\
 \tableline
 & 3.4 & 0.467 & 0.014 \\
 $WISE$ & 4.6 & 0.327 & 0.005  \\
 & 12 & 0.422 & 0.004  \\
 & 22 & 3.33 & 0.02  \\
 %AKARI \\
 \tableline
 & 9 & 0.51 & 0.012 \\
 & 18 & 1.76 & 0.03  \\
 $AKARI$ & 65 & 12.5 & 0.8 \\
 & 90 & 10.9 & 0.6 \\
 & 140 & 5.34 & 1.0  \\
%IRAS \\
 \tableline
 & 12 & 0.605 & 0.0424 \\
 $IRAS$ & 25 & 4.42 & 0.309  \\
 & 60 & 14.1 & 1.41  \\
 & 100 & 9.38 & 0.938 \\
%SCUBA \\
 \tableline
 & 450 & 0.218 & 0.037 \\
 SCUBA& 850 & 0.041 & 0.0024 \\
 & 1100 & 0.039 & 0.013 \\
 & 1350 & 0.0117 & 0.0011 
\enddata
\tablecomments{The EXPORT {\it UBVRI}, \textit{IRAS}, 
               and SCUBA fluxes are taken from 
               Mendigut{\`i}a et al.\ (2012).
               The 2MASS, \textit{WISE}, 
               and \textit{AKARI} photometric
               fluxes are extracted from the online catalogs.
               } 
\end{deluxetable}
%%%%%%%% Table 1: SED %%%%%%%%

\section{Data}\label{sec:data}
We compile all the photometric data for HD\,34700 
available in the literature. The \textit{UBVRI}, 
\textit{IRAS}, and SCUBA fluxes 
are taken from Mendigut{\`i}a et al.\ (2012) 
and references therein. The near- to far-IR fluxes are 
extracted from the 2MASS 
\textit{All-sky Point Source Catalog},\footnote{%
  http://www.ipac.caltech.edu/2mass/releases/allsky
  }
the \textit{WISE} \textit{All-sky Source Catalog},\footnote{%
  http://wise2.ipac.caltech.edu/docs/releases/allsky
  }
the \textit{AKARI Infrared Camera} (IRC)
\textit{All-sky Survey Point Source Catalogue}
(Version 1.0, Ishihara et al.\ 2010), 
and the $AKARI$ \textit{Far Infrared Surveyor} (FIS) 
\textit{All-sky Survey Bright Source Catalogue}
(Version 1.0).\footnote{%
  http://www.ir.isas.jaxa.jp/AKARI/Observation/PSC/Public
  } 
All the photometric 
fluxes and uncertainties reported in the literature
are summarized in Table~\ref{tab:flx}.
For the spectroscopic data of HD\,34700, 
we adopt the $\simali$1.9--4.2$\mum$ \textit{IRTF}/SpeX 
spectrum of Smith et al.\ (2004) and the $\simali$5--14$\mum$
\textit{Spitzer}/IRS spectrum of Sch\"utz et al.\ (2009).
A combination of both photometric and spectroscopic data
enables us to perform a comprehensive modeling 
of the observed spectral energy distribution (SED)
to examine all the major PAH features 
from $\simali$3 to $\simali$14$\mum$
as well as the dust properties simultaneously.

\begin{figure*}[htbp]
\epsscale{1.00}
\plotone{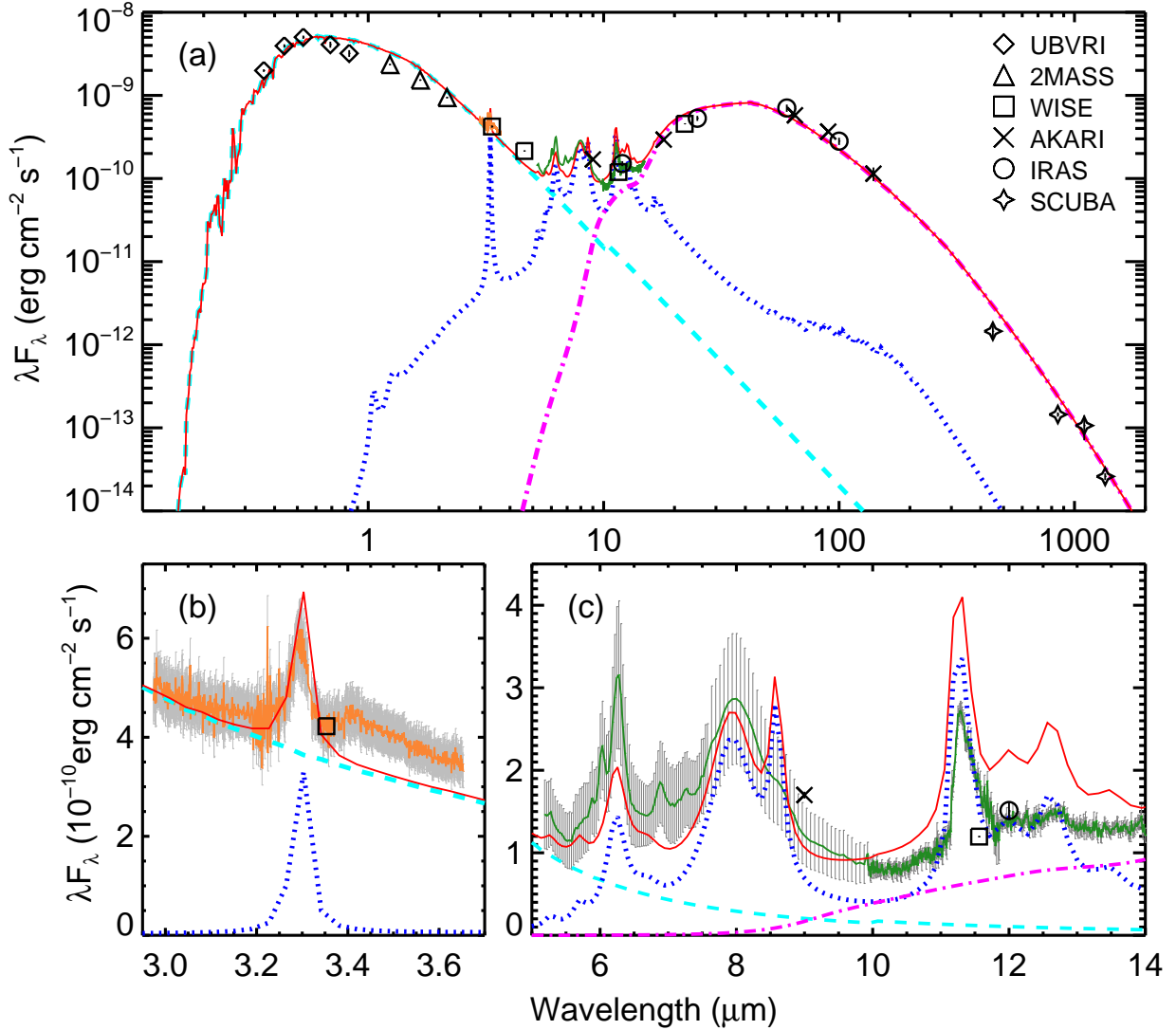}
\caption{\footnotesize
         \label{fig:sed}
         (a) Spectral energy distribution of HD\,34700. 
         The photometric fluxes are shown with symbols 
         (see Table \ref{tab:flx}). 
         The near-IR \textit{IRTF}/SpeX spectrum 
         (Smith et al.\ 2004) and mid-IR \textit{Spitzer}/IRS
         spectrum (Sch\"utz et al.\ 2009) are shown 
         as the orange and green lines, respectively. 
         The photometric uncertainties are overlaid 
         although most of them are smaller than 
         the size of the symbols.
         Our best-fit model spectrum is presented 
         as the red solid line, which is composed of 
         the stellar (cyan dashed line), 
         PAHs (blue dotted line), and 
         dust (magenta dashed--dotted line) components.
         (b)--(c) Same as panel (a), but enlarged views of 
         the \textit{IRTF}/SpeX spectrum
         and the \textit{Spitzer}/IRS spectrum
         of HD\,34700, respectively,
         with measured uncertainties overlaid (gray).
         }
\end{figure*}

Figure~\ref{fig:sed}(a) shows the broadband photometry 
of HD\,34700 together with the \textit{IRTF}/SpeX 
and \textit{Spitzer}/IRS spectra.
In addition, we plot the Kurucz stellar atmospheric 
model spectrum for $\Teff=6000\K$ and a surface gravity 
of $\lg g=4.0$ (Kurucz 1979). 
To correct for the interstellar extinction,
the knowledge of the visual extinction ($A_V$)
toward HD\,34700 is required. In the literature,
a range of $A_V$ values has been reported 
(e.g., $A_V$\,$\approx$\,0--0.68$\magni$,
Coulson et al.\ 1998, van den Ancker 1998, 
Acke \& van den Ancker 2004),
but it is as uncertain as its distance.
In modeling the observed SED, 
we set the distance $d$ 
and visual extinction $A_V$ as free parameters.
We search for the best values that reproduce 
the photometric fluxes with the Kurucz model. 
It is found that $d\approx260\pc$ and $A_V\approx0.5\magni$
closely reproduce the photometric fluxes 
in the ultraviolet/optical and near-IR bands.
Therefore, we adopt $d=260\pc$ with $A_V=0.5\magni$
for further analysis in this work.\footnote{%
  If $A_V=0$, the same Kurucz model with $d\approx320\pc$
  can also closely reproduce the photometric fluxes 
  up to the {\it J, H, K$_{\rm s}$} bands 
  but is deficient in the \textit{WISE} 3.4 and
  4.6$\mum$ bands. If this true, the excess emission
  at the \textit{WISE} 3.4 and 4.6$\mum$ bands
  may indicate the existence of
  an inner, warm ``zodiacal cloud'' 
  in the HD\,34700 disk.
  }

Figures~\ref{fig:sed}(b),(c) focus on 
the 3.3$\mum$ PAH feature and the PAH features 
at longer wavelengths, respectively.
It is interesting to note that the 3.3$\mum$ 
aromatic C--H stretch feature of HD\,34700
is accompanied by a weaker feature at 3.43$\mum$ 
(see Figure~\ref{fig:sed}(b)),
which is generally ascribed to 
aliphatic C--H stretch 
(Joblin et al.\ 1996, Li \& Draine 2012, 
Yang et al.\ 2013).
In addition to the 3.43$\mum$ C--H stretch,
aliphatic hydrocarbon materials also have 
two C--H deformation bands
(see Li \& Draine 2012). 
The \textit{Spitzer}/IRS spectrum of HD\,34700
reveals the presence of these two features 
at $\simali$6.89 and $\simali$7.23$\mum$
(see Figure~\ref{fig:sed}(c)).
Compared with the PAH features seen in 
the diffuse ISM (e.g., see Tielens 2008), 
the 7.7$\mum$ feature is somewhat red-shifted
and appreciably broadened, 
while the 8.6$\mum$ feature is much weaker.
These characteristics belong to either class B 
or C following the PAH spectral classification of 
Peeters et al.\ (2002), which is typically found 
in planetary nebulae and isolated Herbig Ae/Be stars. 

%%%%%%%%%%%%%%% Figure 1: SED %%%%%%%%%%%%%%%
%%%%%%%%%%%%%%% Figure 1: SED %%%%%%%%%%%%%%%

\section{Model}\label{sec:model}
To simultaneously model the dust emission 
and the \textit{IRTF}/SpeX 
and the \textit{Spitzer}/IRS spectra,
we adopt the porous dust model of 
Li \& Lunine (2003a)
and the astro-PAH moel of Li \& Draine (2001)
and Draine \& Li (2007).

\subsection{PAHs}\label{sec:pah}
Following Li \& Draine (2001) 
and Li \& Lunine (2003a),
we assume a log-normal size distribution 
for the PAHs in the HD\,34700 disk.
The log-normal size distribution function
is characterized by two parameters, 
$a_{0}$ and $\sigma$,
which determine 
the peak location and the width of 
the log-normal distribution, respectively:
\begin{equation}\label{eq:dnda_pah}
dn_{\rm PAH}/da = 
\frac{1}{\sqrt{\pi/2}\,\sigma
        \left\{ 1 - {\rm erf}\left[
        \ln\left(a^{\rm PAH}_{\rm min}/a_0\right)/\sqrt{2}\sigma
        \right] \right\}}     
        \frac{1}{a} \exp \left\{ - \frac{1}{2} 
       \left[ \frac{\ln (a/a_{0})}{\sigma} \right]^2 \right\},
       ~~{\rm for}\  a > \apahmin ~.
\end{equation}
We set the lower cutoff of the PAH size
as $a^{\rm PAH}_{\rm min}\equiv3.5\Angstrom$ 
corresponding to $\sim20$ carbon atoms in a PAH
molecule (i.e., $N_{\rm C}\approx20$), 
which is the minimum survival size for PAHs 
in the diffuse ISM (see Li \& Draine 2001).
We treat $a_0$ and $\sigma$ as free parameters.
The total PAH mass will be derived from
fitting the observed PAH emission features.

For the PAH absorption cross sections, 
we adopt those of Draine \& Li (2007).
Two types of PAHs are distinguished:
neutral PAHs and charged PAHs.
For charged PAHs, we do not distinguish 
either cations from anions
or multiply charged PAHs
from singly charged PAHs. 

The astro-PAH model of Li \& Draine (2001)
and Draine \& Li (2007) represents 
the absorption cross section of each PAH feature 
by a Drude profile, which is expected for classical 
damped harmonic oscillators (see Li 2009a).
Drude profiles are characterized 
by their peak wavelengths, widths, and strengths 
(i.e., the wavelength-integrated absorption 
cross sections).
As mentioned earlier (see Section \ref{sec:data}),
the PAH feature profiles detected in HD\,34700
somewhat differ from those typically seen in 
the diffuse ISM. In modeling the PAH emission
features of HD\,34700, we slightly adjust 
the peak wavelength and width for several features,
but with the strength fixed. We note that it is 
not unphysical to adjust the peak wavelength and width 
as long as the strength is unaltered.

We will only model the major PAH features 
such as those at 3.3, 6.2, 7.7, 8.6 and 11.3$\mum$.
Nevertheless, the intensities of the minor, 
aliphatic C--H features at 3.43, 6.89, and 7.23$\mum$ 
are measured and will be briefly discussed 
in Section \ref{sec:pah}.
The aliphatic C--H features
of HD\,34700 and other disks
will be fully investigated in 
a separate paper 
(J. Y.~Seok \& A.~Li 2015, in preparation).

The PAH features shown in Figure~\ref{fig:sed} 
manifest that both neutral and ionized PAHs
exist in the HD\,34700 disk
because ionized {\small PAHs} emit strongly 
at 6.2, 7.7, and 8.6$\mum$,
in contrast, neutral {\small PAHs} emit 
strongly at 3.3 and 11.3$\mum$.
The charge distribution of PAHs is determined by 
the balance between the photoionization 
and the electron recombination 
(Bakes \& Tielens 1994, Weingartner \& Draine 2001),
which is proportional to 
$U\sqrt{T_{\rm gas}}/n_{\rm e}$,
where $U$ is the ultraviolet (UV) starlight 
intensity, $T_{\rm gas}$ is the gas temperature,
and $n_{\rm e}$ is the electron density.
Moreover, the PAH size also affects its charging.
Following the approaches described in Li \& Lunine (2003a),
we calculate the photoionization rate ($k_{\rm ion}$) 
and the electron recombination rate ($k_{\rm rec}$)
for PAHs of a given size at a given distance ($r$) from
the central star (see Appendices A and B in Li \& Lunine 2003a).

%%%%%%%%%%%%%%% Figure 2: Rad. profiles of ne, Tgas, U  %%%%%%%%%
\begin{figure}[tbp]
\epsscale{0.50}
\plotone{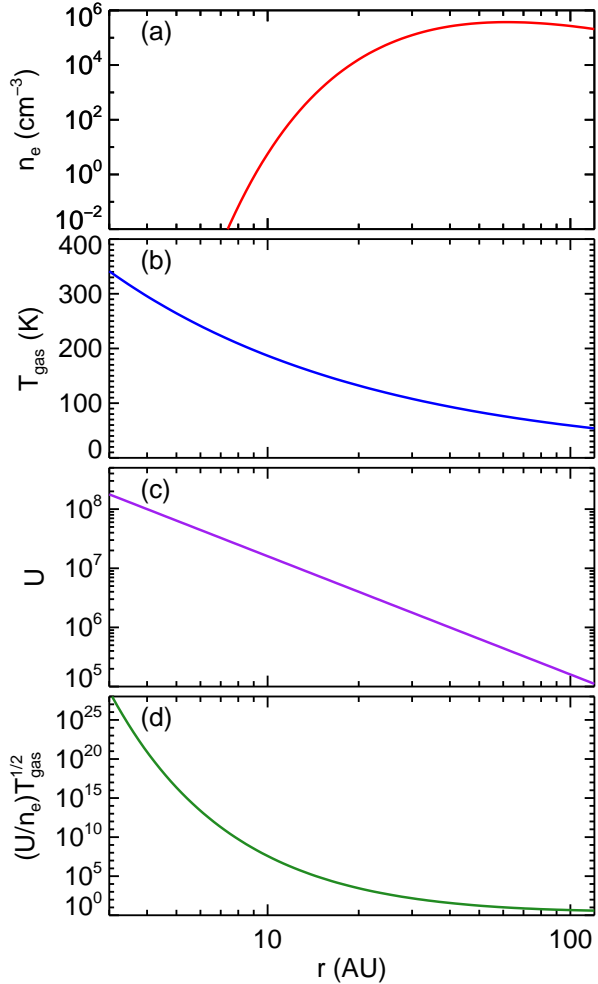}
\caption{\footnotesize
         \label{fig:ne}
         Radial profiles of 
         (a) the electron density, $n_{\rm e}$ (cm$^{-3}$),
         (b) the gas temperature, $T_{\rm gas}$ (K),
         (c) the starlight intensity, $U$,
             in unit of the 912$\Angstrom$--1$\mum$
             MMP83 ISRF (e.g., see Equation~(1) of
             Li \& Lunine 2003a), and
         (d) the ionization parameter 
             $U\sqrt{T_{\rm gas}}/n_{\rm e}$ 
         as a function of distance, $r$ (AU),
         from the central star in the HD\,34700 disk. 
         Since the ionization balance of PAHs is controlled 
         by $U\sqrt{T_{\rm gas}}/n_{\rm e}$ 
         (Bakes \& Tielens 1994, Weingartner \& Draine 2001),
         the trend seen in panel (d) depicts the overall charging
         of PAHs depending on the distance to the central star. 
         }
\end{figure}
%%%%%%%%%%%%%%% Figure 2: Rad. profiles of ne, Tgas, U %%%%%%%%%

%%%%%%%%%%%%%%% Figure 3: PAH ioni./recomb. %%%%%%%%%%%%%%%
\begin{figure}[tbp]
\epsscale{0.50}
\plotone{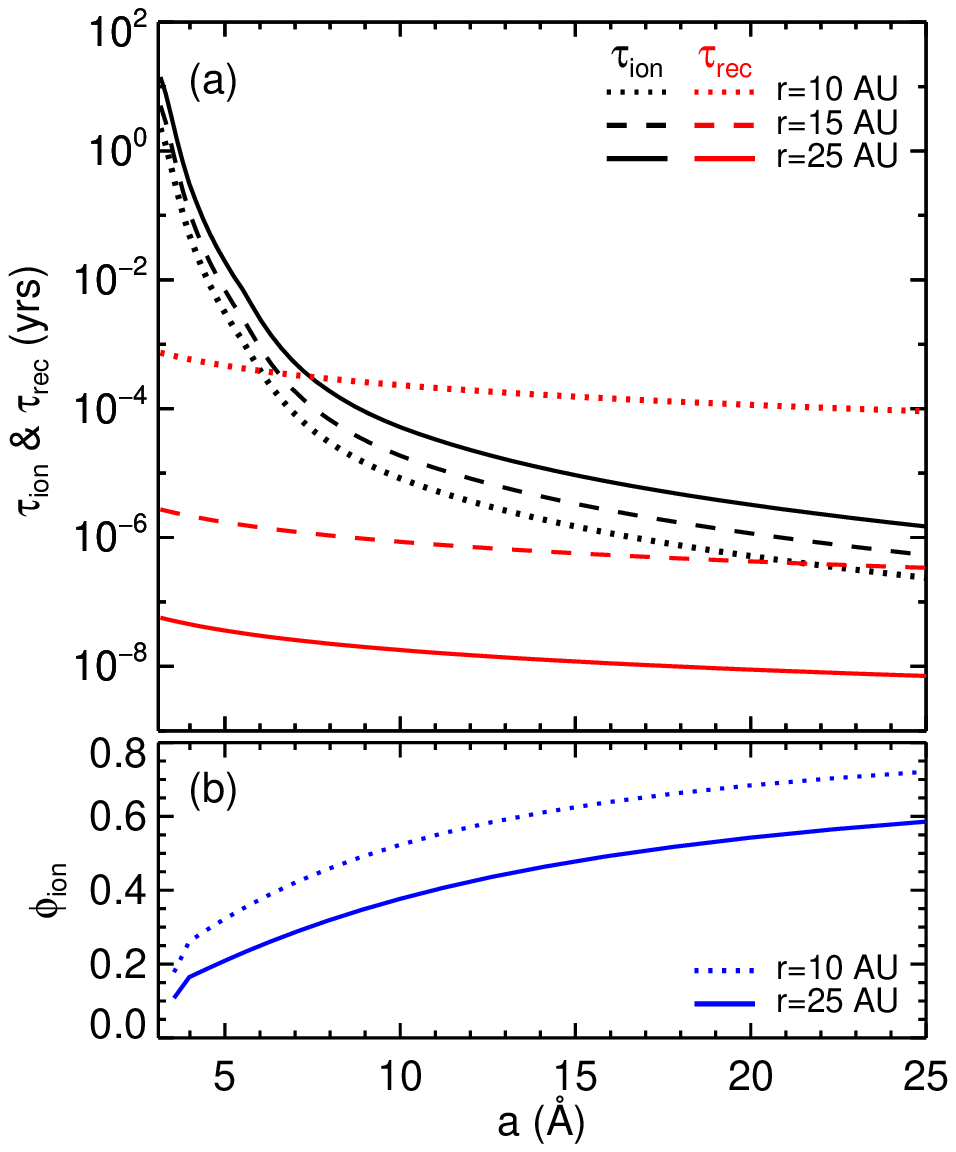}
\caption{\footnotesize
         \label{fig:ion}
         Upper panel (a) Photoionization ($\tau_{\rm ion}$) 
             and electron recombination ($\tau_{\rm rec}$)
             timescales for PAHs in the size range of 
             $3.5\Angstrom<a<25\Angstrom$ 
             at $r=10\AU$ (dotted lines), 
             $15\AU$ (dashed lines), and
             $25\AU$ (solid lines) 
             from the central star.
             In calculating $\tau_{\rm rec}$, 
             we estimate $n_{\rm e}$ from $M({\rm H_2})$,
             the mass of H$_2$ determined by
             Zuckerman et al.\ (1995),
             assuming that the cosmic-ray ionization of H$_2$
             is the dominant contributor to the electrons 
             in the HD\,34700 disk.
        Lower panel (b) ionization fraction 
             ($\phi_{\rm ion}$) of PAHs 
             as a function of size
             at $r=${\bf10 and} 25$\AU$.
             }
\end{figure}
%%%%%%%%%%%%%%% Figure 3: PAH ioni./recomb. %%%%%%%%%%%%%%%

To calculate $k_{\rm rec}$, we need to know 
the electron density $n_{\rm e}$ 
as a function of $r$, 
the distance from the central star.
We assume the cosmic-ray ionization 
of H$_2$ to be the major source of the electrons 
in the HD\,34700 disk.
The electron density is derived from 
$n_{\rm e}\approx\tau_\star \varsigma_{\rm CR}n_{\rm H_2}$, 
where $\varsigma_{\rm CR}\approx3\times10^{-17}\s^{-1}$ is 
the cosmic ionization rate and 
$\tau_\star\approx10\Myr$ is the stellar age. 
We obtain $n_{\rm H_2}$ from $M({\rm H_2})$,
the total mass of H$_2$ 
in the HD\,34700 disk.
We adopt $M({\rm H_2}) = 1502\,\Mearth$
after scaling the original value of
$M({\rm H_2})\approx80\,\Mearth$ 
at $d\approx60\pc$ of Zuckerman et al.\ (1995)  
to $d\approx260\pc$,
the distance adopted in this work.
Assuming a vertical hydrostatic equilibrium
for the disk (see Equations (B2)--(B4)
in Li \& Lunine 2003a), we derive $n_{\rm H_2}(r)$, 
the spatial distribution of H$_2$,
and then $n_{\rm e}$ (see Figure~\ref{fig:ne}(a)).
Also, we take 
$T_{\rm gas}(r)\approx\left(R_\star/2r\right)^{1/2}\Teff
\approx591\left(r/{\rm AU}\right)^{-1/2}\K$,
where $R_\star\approx0.0194\AU$ 
is the stellar radius 
(see Figure \ref{fig:ne}(b)).
%stellar mass of $M_\ast\approx1.2~m_\sun$.
In Figure \ref{fig:ne}(d),
we show the ionization parameter
$U\sqrt{T_{\rm gas}}/n_{\rm e}$ as a function
of the distance $r$ from the central star,
where the starlight intensity, $U$,
is in unit of the 912$\Angstrom$--1$\mum$
local interstellar radiation field of
Mathis et al. (1983; MMP83).

In Figure~\ref{fig:ion}(a), 
the photoionization timescales 
($\tau_{\rm ion}\equiv1/k_{\rm ion}$)
and the electron recombination timescales 
($\tau_{\rm rec}\equiv1/k_{\rm rec}$) 
as a function of PAH size are depicted 
for given distances from the central star
($r=10$, 15, and 25$\AU$). 
Note that $r=25\AU$ is considered to 
be the cutoff distance 
of the PAH distribution in the HD\,34700 disk
since at $r>25\AU$ the dust temperatures
drop to $T<120\K$ (see Section \ref{sec:results}),
so the dust willx acquire an ice mantle 
and PAHs will condense onto the ice mantle.
As discussed in Li \& Lunine (2003a),
$\tau_{\rm ion}$ decreases for larger PAHs 
since the ionization threshold is inversely 
proportional to the PAH size. 
Also, $\tau_{\rm ion}$ decreases as PAHs get closer 
to the central star. This is because the starlight 
intensity is enhanced (i.e., $U\propto r^{-2}$,
see Figure \ref{fig:ne}(c)).

In comparison with $\tau_{\rm ion}$, 
$\tau_{\rm rec}$ is found to be much shorter 
than $\tau_{\rm ion}$ at $r=25\AU$
whereas $\tau_{\rm rec}$ becomes comparable to 
$\tau_{\rm ion}$ at $r\simeq15\AU$ for large PAHs. 
At $r=10\AU$, PAHs with $a\ga6\Angstrom$ 
have $\tau_{\rm ion}$ even shorter than $\tau_{\rm rec}$. 
This indicates that PAHs, in particular large PAHs, 
can attain an ionized state in the inner region 
while PAHs in the outer region are more likely to 
be neutral (or negatively charged).

Using the photoelectric emission 
and electron capture rates of 
Weingartner \& Draine (2001),
we investigate the PAH charging process
in the HD\,34700 disk.
At a given distance from the star,
we calculate the ionization fraction 
($\phi_{\rm ion}$) of PAHs as a function 
of PAH size, 
which is the probability 
of finding a PAH molecule in a nonzero charge state.
As shown in Figure~\ref{fig:ion}(b)
for $\phi_{\rm ion}(a)$ at $r=10$ and $25\AU$,
small PAHs are more likely to be neutral, 
and large PAHs are more likely to be ionized. 
This is consistent with the comparison 
between the photoionization and electron recombination 
timescales (see Figure~\ref{fig:ion}(a)).
Overall, $\phi_{\rm ion}$ decreases
as the distance $r$ increases, consistent 
with the variation trend of 
$U\sqrt{T_{\rm gas}}/n_{\rm e}$ with $r$
(see Figure \ref{fig:ne}(d)) 
which determines the charging of PAHs 
as a function of $r$. 
Roughly speaking, PAHs near the central star are
fully ionized since $\tau_{\rm ion}\ll \tau_{\rm rec}$
even for the smallest PAHs.
At $\simali$15--25$\AU$, however,
because of $\tau_{\rm ion}\gtsim \tau_{\rm rec}$
for PAHs of $a\simlt20\Angstrom$
(see Figure~\ref{fig:ion}(a)),
a fraction of them will be negatively charged,
so $\phi_{\rm ion}$ will be nonzero
even for the smallest PAHs at $r=25\AU$
(see Figure~\ref{fig:ion}(b)).
Therefore, we expect a mixture of 
neutral and charged PAHs (cations or anions) 
in the HD\,34700 disk. 
Note that, following Li \& Draine (2001), 
we do not distinguish cations and anions 
when we model the IR emission of PAHs.

\subsection{Dust}\label{sec:dust}
The porous grain model of Li \& Lunine (2003a) 
for protoplanetary and debris disks
assumes that the dust consists of
porous aggregates of amorphous silicate 
and carbonaceous materials,
and the dust in regions cooler than $\simali$120$\K$
is further coated with H$_2$O-dominated ices.
We take the dielectric functions of
``astronomical'' silicate of Draine \& Lee (1984)
for the silicate component,
amorphous carbon of Rouleau \& Martin (1991)
for the carbonaceous component, and
``dirty'' ice of Li \& Greenberg (1998)
for the ice component.
Assuming a spherical shape,
the absorption cross sections of porous dust
are calculated from Mie theory combined with
the Maxwell--Garnett and Bruggeman 
effective medium theories (see Li 2009a).
We adopt the mass mixing ratios of
Li \& Lunine (2003b) derived from 
cosmic abundance considerations.
We first employ the Maxwell-Garnett 
effective medium theory 
to calculate the average dielectric functions 
for the ice-coated silicate subgrains 
and the ice-coated carbonaceous subgrains
(see Equations~(7),\,(8) of Li \& Lunine 2003a)
and then use the Bruggman effective medium theory 
to calculate the average dielectric functions 
for the porous dust aggregate
(Equation~(9) of Li \& Lunine 2003a).
The porosity (or fluffiness) of a grain 
is quantified by the fractional volume 
of vacuum ($P$).
Following Li \& Lunine (2003a, 2003b),
we adopt $P=0.90$ for porous dust 
and $P^{\prime}\approx0.73$ for 
the ice-coated porous dust, 
and the choice of $P$ 
will be discussed 
in Section~\ref{sec:results}.

For the spherical dust,
we adopt a power-law size distribution: 
$dn/da\propto a^{-\alpha}$,
where $a$ is the spherical radius of the dust,
and $\alpha$ is the power-law index. 
We take the lower cutoff size to be $\amin=1\mum$
and the upper cutoff size to be $\amax=1\cm$. 
Li \& Lunine (2003a) have shown that changing 
$\amin$ and $\amax$ does not 
significantly alter the model-fitting.
This will be discussed 
in Section~\ref{sec:results}.
We vary $\alpha$ from 0.1 to 4.0 to 
search for the best fit.

For the spatial (radial) distribution of 
the dust in the HD\,34700 disk,
instead of using a simple power-law 
with sharp inner and outer boundary cutoffs,
we propose a more physical formula for 
the spatial distribution, 
which is defined as
\begin{equation}
\label{eq:dndr}
\frac{dn}{dr}\propto\left(1-\frac{r_{\rm min}}{r}\right)^\beta
\left(\frac{r_{\rm min}}{r}\right)^\gamma~,
\end{equation}
where $r_{\rm min}$ is the inner boundary of the disk,
which is set to be the distance where the dust temperature 
reaches $\simali$1500$\K$ and dust sublimates.
This functional form has the advantage that on one hand,
it resembles a power-law $dn/dr \propto r^{-\gamma}$ at
larger distances ($r\gg\rmin$), and on the other hand
it peaks at $\rp = \rmin \left(\beta+\gamma\right)/\gamma$,
unlike the simple power-law which peaks at $\rmin$. 
The latter is unphysical as it assumes that the dust 
piles up at $\rmin$ where dust actually sublimates. 
%
%\begin{equation}
%\beta=\gamma\times(r_{\rm peak}/r_{\rm min}-1),
%\label{eq:beta}
%\end{equation} 
%
We take $\rmin=0.3\AU$ since the equilibrium temperature 
of micrometer-sized dust reaches $\simali$1500$\K$ 
at $r$\,$\simali$0.3$\AU$ from the central star.
We take the outer disk boundary to be $\rmax=1000\AU$.
Our model is not sensitive to $\rmax$,
but $\rmax=1000\AU$ is a reasonable assumption 
in view of the distance to the nearest companion star 
%(beside the spectroscopic binary) 
in the system of HD\,34700.\footnote{%
  The angular distance from the central star
  to the stellar component B 
  is $\simali$5\farcs2 (Sterzik et al.\ 2005), 
  which corresponds to $\simali$1350$\AU$
  at a distance of $d=260\pc$.
  }  
In summary, we have two free parameters,
$r_{\rm p}$ and $\gamma$,
for the dust spatial distribution.

%%% Results %%%
\section{Results}\label{sec:results}
Figure~\ref{fig:sed} shows our best-fit model 
that reasonably reproduces all the observational data.
The best-fit model includes emission from 
a mixture of neutral/ionized PAHs
and porous dust with a power-law size distribution 
index of $\alpha\approx3.5$ and
a total dust mass of $M_{\rm dust}\approx21.9\Mearth$. 
The disk geometry is characterized 
with $r_{\rm p}\approx100\AU$ 
and $\gamma\approx2$
(see Equation\,(\ref{eq:dndr})).
The PAH component is determined by 
$a_0\approx3.5\Angstrom$ and $\sigma\approx0.3$, 
and the total PAH mass is 
$M_{\rm PAH}\approx 4.53\times10^{-7}
\left(r/{\rm AU}\right)^2\Mearth$ 
if all PAHs are at a distance of $r$ 
from the central star. 
We do not have any prior information how the PAHs are
spatially distributed in the HD\,34700 disk.
Due to their stochastic heating nature
(Draine \& Li 2001), the PAH emission
spectral profiles do not vary with 
the starlight intensity. Therefore,
the PAH emission features should 
essentially remain identical for PAHs
at different distances from the central star,
except that the flux level scales with $r^{-2}$. 
Consequently, the PAH mass required to account for 
the observed PAH emission features scales with $r^2$.

%%%%%%%%%%%%%%% Figure 4: dn/da for PAHs %%%%%%%%%%%%%%%
\begin{figure}[tbp]
\epsscale{0.50}
\plotone{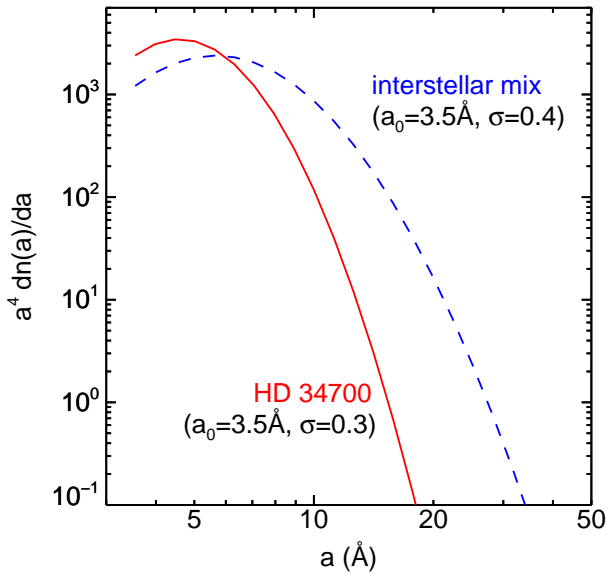}
\caption{\footnotesize
         \label{fig:pahsz}
         Log-normal size distribution for 
         the PAH population. 
         The best-fit model with $a_0=3.5\Angstrom$
         and $\sigma=0.3$ is compared with that of 
         the interstellar PAH model
         ($a_0\approx3.5\Angstrom$ 
          and $\sigma\approx0.4$, Li \& Draine 2001). 
          The size distribution is expressed by multiplying 
          $a^4$ to show the mass distribution per logarithmic 
          PAH radius.
          }
\end{figure}
%%%%%%%%%%%%%%% Figure 4: dn/da for PAHs %%%%%%%%%%%%%%%

The PAH size distribution with the best fit parameters 
is shown in Figure~\ref{fig:pahsz} together with that of 
the interstellar mixture (i.e., $a_0\approx3.5\Angstrom$
and $\sigma\approx0.4$, Li \& Draine 2001). 
Compared with the diffuse ISM, the PAH emission features
observed in the HD\,34700 disk are in favor of smaller PAHs.

%%%%%%%%%%%%%%% Figure 5: T(dust) %%%%%%%%%%%%%%%
\begin{figure}[tbp]
\epsscale{1.1}
\plottwo{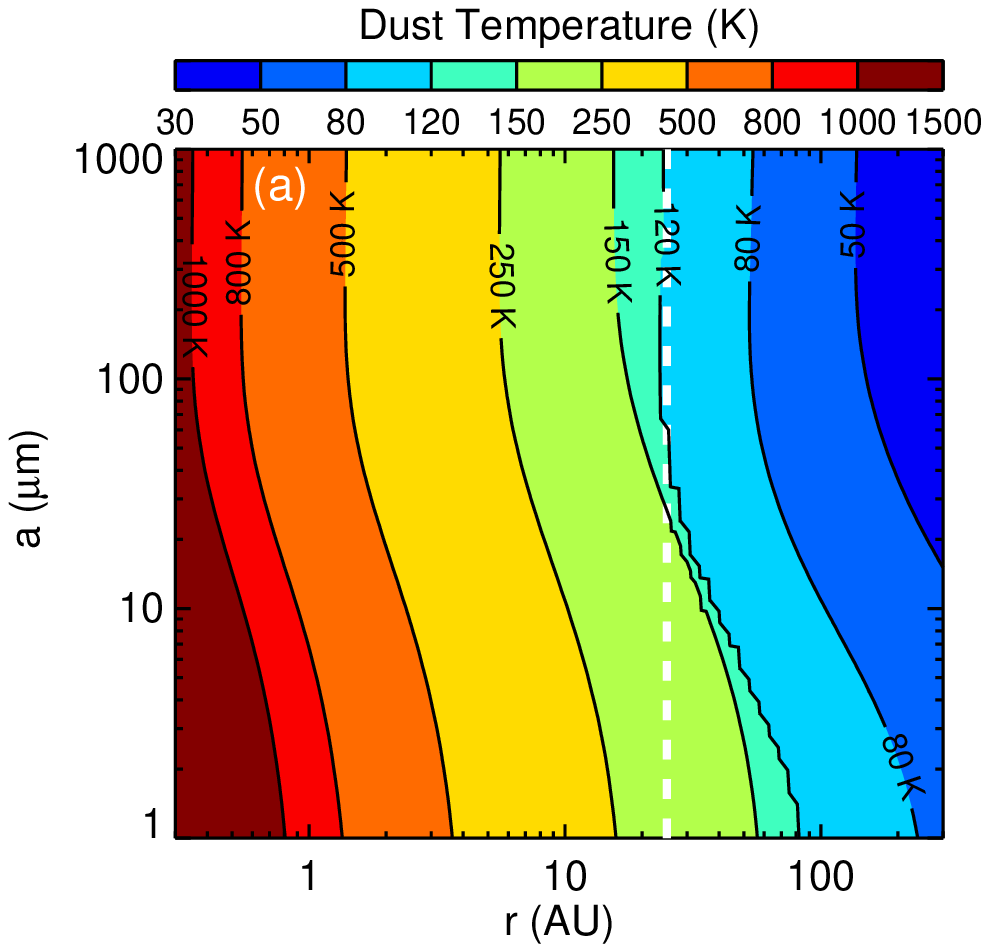}{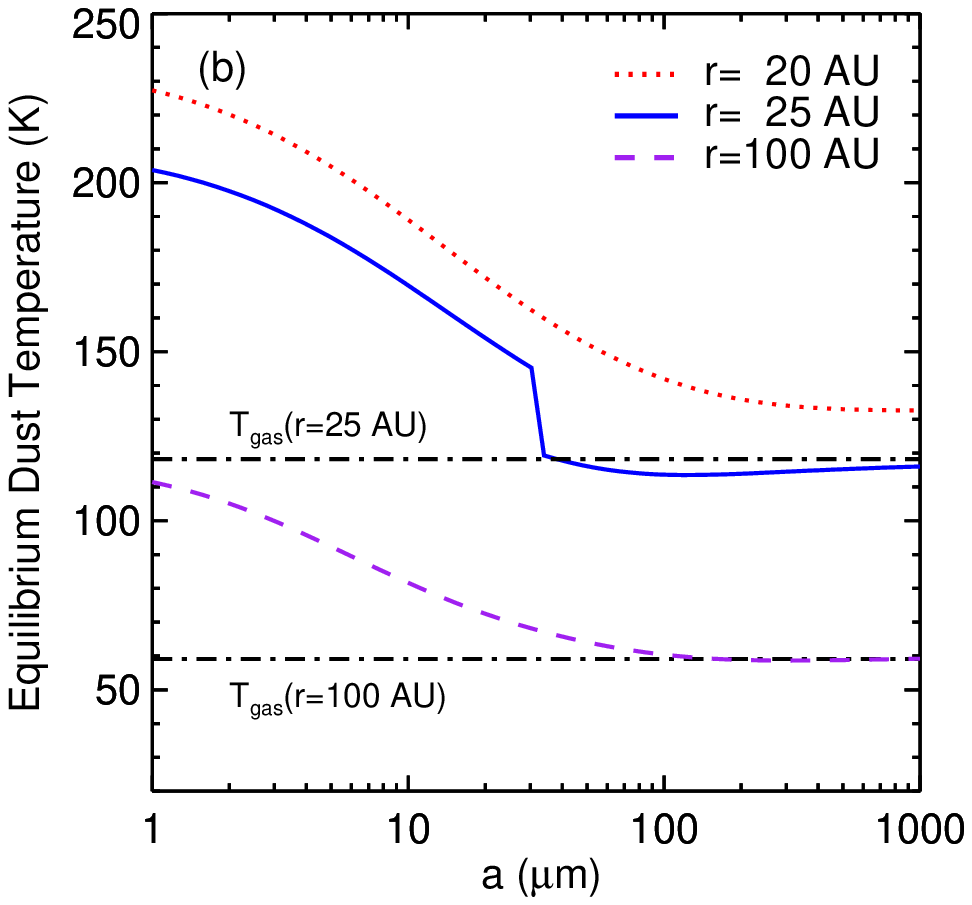}
\caption{\footnotesize
         \label{fig:Teq}
         (a) Temperatures of porous dust
         (of porosity $P=0.90$ for $T>120\K$
         and $P^{\prime}=0.73$ for $T<120\K$)  
         as a function of grain size ($a$) 
         and radial distance ($r$)
         from the central star.
         (b) Temperatures of porous dust 
         as a function of grain size 
         at $r=20\AU$ (dotted lines),
         $r=25\AU$ (solid lines), and
         $r=100\AU$ (dashed lines).
         A sudden temperature drop around $a\approx30\mum$ 
         is caused by the ice sublimation at $T\approx120\K$. 
	 For comparison, the gas temperatures 
         $T_{\rm gas}\approx591\left(r/{\rm AU}\right)^{-1/2}\K$ 
         at the same distances are overlaid. 
         }
\end{figure}
%%%%%%%%%%%%%%% Figure 5: T(dust) %%%%%%%%%%%%%%%

Our model yields a total dust IR flux of 
$\int_{912\Angstrom}^{\infty} F_\lambda d\lambda
\approx1.31\times10^{-9}\erg\cm^{-2}\s^{-1}$, 
of which the PAH component contributes 
$\simali$10\%.\footnote{%
  In contrast, in the Galactic diffuse ISM
  the PAH component accounts for $\simali$1/3
  of the total IR luminosity.
  } 
This results in a total IR luminosity 
of $L_{\rm IR}\approx2.75\Lsun$ 
at $d=260\pc$ and a ratio of the IR luminosity 
to the stellar luminosity of 
$L_{\rm IR}/L_\star\approx0.13$.
%(for HD\,34700, $L_\star\approx20.4\Lsun$). 
Previous studies estimated 
$L_{\rm IR}/L_\star$\,$\simali$0.14 
(Sheret et al.\ 2004)
or $\simali$0.15 (Sylvester et al.\ 1996),
in close agreement with our estimate.

The best-fit model predicts that the peak radius 
of the dust spatial distribution in the HD\,34700 disk 
is located at $\rp\approx100\AU$ from the central star. 
The peak radius $\rp$ derived here can be
compared with the inner radius 
(i.e., $r_{\rm min}$) of the disk models 
proposed in the literature
which assume a simple power-law
dust spatial distribution 
%between the inner and outer radii
and a peak distribution at $\rmin$.
Sylvester et al.\ (1996) and Dent et al.\ (2005) derived 
$r_{\rm min}$\,$\approx$\,24--46$\AU$ (at $d\approx55\pc$) 
and $\lesssim30\AU$ (at $d\approx125\pc$), respectively.
If we scale them to those at $d=260\pc$, 
our estimate is consistent with the former
but is slightly larger than the latter.

Using the power-law size distribution 
and the spatial distribution of the dust
(see Equation (\ref{eq:dndr})), 
we estimate the total dust mass in the disk
to be $M_{\rm dust}\approx21.9\Mearth$. 
There have been various estimates of
the dust mass of the HD\,34700 disk 
(e.g., see Zuckerman et al.\ 1995, 
Coulson et al.\ 1998, Sheret et al.\ 2004, 
Dent et al.\ 2005). 
Although these previous studies adopted 
different distances to HD\,34700,
ranging from $d\approx60\pc$ to $d\ga180\pc$,
the derived dust mass ranges from
$M_{\rm dust}\approx7$
to $\simali$30$\Mearth$ after 
scaling to $d=260\pc$ 
(i.e., $M_{\rm dust}\propto d^2$), 
which overlaps with our estimate.

We calculate the equilibrium
temperature, $T(r, a)$, of porous dust 
as a function of the radial distance $r$ 
from the star and grain size $a$ 
(see Figure \ref{fig:Teq}(a)).
The temperature decreases 
with increasing grain size\footnote{%
  For large grains with $a\gtsim100\mum$,
  they emit like a blackbody and their temperatures
  are independent of grain size.
  }
and distance from the star.
Figure~\ref{fig:Teq}(b) shows the equilibrium 
dust temperatures as a function of dust size
at specific distances ($r=20$, 25, and 100$\AU$).
At $r=25\AU$, we see a sudden temperature drop 
around $a\approx30\mum$. 
This is caused by the ice sublimation 
at $T\approx120\K$:
porous dust larger than $a$\,$\simali$30$\mum$
attains $T\approx120\K$ at $r\approx25\AU$.
When the silicate and carbonaceous subgrains
acquire an ice mantle at $T<120\K$,
their aggregates will be less absorptive 
and therefore attain 
a lower equilibrium temperature.

The location where dust grains have low enough 
temperatures to capture volatile molecules
to form ice mantles is called the ``snowline.''
The snowline plays an important role 
in the PAH chemistry
since free-flying PAHs will 
condense onto the ice mantles. 
As shown in Figure \ref{fig:Teq}(a),
the transition from non-icy grains 
to icy grains occurs at different distances
from the star for grains of different sizes:
for smaller grains, the ``snowline'' is further
away from the star, 
and by $r\approx80\AU$
all grains become ice-coated.
For simplicity, in the following, we adopt $r=25\AU$ as 
the snowline
of the HD\,34700 disk.

Although the observed SED is closely reproduced
by our best-fit model, there could exist some 
flexibilities for the model parameters.
We briefly describe here the robustness of our model
and refer to Li \& Lunine (2003a), 
who explicitly discussed
all the parameters of 
the porous dust model for disks
(see their Section 4.1 for details).
First, we examine the adopted dust parameters
$\amin$, $\amax$, and $P$
while other parameters remain unchanged.
We differ $\amin$ between $0.1\mum$ and $10\mum$.
While models with a large $\amin$ ($\simali$10$\mum$)
emit too little in the mid-IR at $\lambda\la40\mum$,
those with a small $\amin$ ($\simali$0.1$\mum$)
deviate the fit at the mid- and far-IR.
Similarly, by varying $\amax$ from $100\mum$ up to $1\cm$,
we find that sufficiently large grains ($\ga0.1\cm$) 
are necessary for reproducing the observed submm emission.
The models with $1000\mum\la\amax\la1\cm$ produce 
almost identical SEDs.
As $\amax$ decreases,
the total dust mass ($M_{\rm dust}$) decreases
by a factor of $\simali$3 at most.
The porosity $P$ is degenerate 
with $a_{\rm min}$ and $\rmin$
since decreasing $a_{\rm min}$ or $r_{\rm min}$ 
or increasing $P$ has the same effect of 
increasing the dust temperature.  
However, a porosity in the range of 
$0.80\simlt P\simlt 0.90$
is expected for dust aggregates 
formed through coagulation
as demonstrated both theoretically 
(Cameron \& Schneck 1965) 
and experimentally (Blum \& Wurm 2008).
Moreover, a porosity of $P\simeq 0.90$ 
is consistent with the mean mass density 
of cometary nuclei 
for which the ice-coated dust aggregates 
are plausible building blocks 
(see Greenberg \& Li 1999).
It is worth noting that
the fluxes calculated 
from models with $0.8\la P \la 0.95$
for the HD\,34700 disk
agree with each other within $\la25\%$.

We also examine 
the dust spatial distribution parameters 
$\rp$ and $\gamma$ 
which are not readily constrained  
due to the lack of spatial resolution
of the disk. 
We find that models with small $\rp$ 
($\simlt80\AU$)
would emit too much around $\lambda\sim10\mum$ 
due to the silicate emission feature,
while models with large $\rp$ ($\ga200\AU$)
fail to explain the far-IR/submm emission.
The $80\AU\la\rp\la150\AU$ models produce
reasonable fits, with $M_{\rm dust}$ varying 
by a factor of $\simali$2 at most.
We also find that varying $\gamma$ 
($1.0\la\gamma\la3.5$; combined with $\rp$) 
does not affect the model-fit significantly.
In summary, 
although different combinations of model parameters 
may be able to explain the observed SED, 
the underlyling results do not change. 
The estimate of $M_{\rm dust}$
is most vulnerable 
(by a factor of $\simali$3 at most),
while the other estimates 
(e.g., $\dot{M}_{\rm RP}$
and $\dot{M}_{\rm PR}$ 
in Section~\ref{sec:rppr}) 
vary within $\simali$20\%.

%%%%%%%%%%%%%%%%%%%%%%%%%%%%%%%%%%
\section{Discussion}\label{sec:discussion}
\subsection{PAHs in the HD 34700 Disk}\label{sec:pah}
As mentioned in Section \ref{sec:results},
due to the single-photon heating nature of PAHs,
the total mass of PAHs in the disk ($M_{\rm PAH}$) 
required to reproduce the observed PAH emission is 
simply scaled by $U$, the starlight intensity.
However, we cannot constrain $U$
(which is inverse proportional to $r^2$,
i.e., $U(r)\propto(r/{\rm AU})^{-2}$)
since the HD 34700 disk has not been 
spatially resolved and hence we do not have any 
prior knowledge of the spatial distribution of PAHs.
One may derive $M_{\rm PAH}\approx2.83\times10^{-4}\Mearth$
by locating all PAHs at $r=25\AU$,\footnote{%
   This means that most of the PAHs are concentrated 
   within $r\sim25\AU$ corresponding to an angular
   size of $\theta\approx0\farcs1$ at a distance 
   of $d\sim260\pc$.
   Future IR observations with a sub-arcsecond 
   spatial resolution such as 
   the \textit{Near Infrared Camera} (NIRCam)  
   onboard the \textit{James Webb Space Telescope} 
   will shed light on the spatial distribution 
   of dust and PAHs in debris disks.
   } 
the location of the snowline in the HD\,34700 disk.
Beyond the snowline, grains
start to be ice-coated (see Figure~\ref{fig:Teq}), 
and the free-flying PAHs will 
condense onto the ice mantles
and will not be excited to emit 
at the observed PAH features.

Even with 
$M_{\rm PAH}\approx2.83\times10^{-4}\Mearth$
for $r=25\AU$,
the PAH-to-dust mass ratio is only 
$M_{\rm PAH}/M_{\rm dust}\approx1.3\times10^{-5}$,\footnote{%
  In HD 34700, the fractional IR luminosity 
  ($L_{\rm PAH}/L_{\rm IR}\sim10\%$, 
   see Section \ref{sec:results})
  emitted by PAHs is much higher than 
  the PAH-to-dust mass ratio.
  This is because PAHs are closer to
  the central star and also much more
  absorptive than the bulk porous dust.
  }
much smaller than that of the Galactic diffuse ISM
($M_{\rm PAH}/M_{\rm dust}\approx0.05$, 
Li \& Draine 2001, Draine \& Li 2007).
The deficit of PAHs relative to dust
has previously been reported 
for debris disks
(e.g., see Li \& Lunine 2003a, Thi et al.\ 2013).
However, the derived PAH-to-dust mass ratio
may not reflect the actual PAH-to-dust mass ratio
in the HD\,34700 disk as there could be a substantial
amount of PAHs in the disk beyond the snowline. 

The minor features at 3.43, 6.89, and 7.23$\mum$ 
shown in the HD\,34700 disk reveal that the PAH
molecules very likely have aliphatic sidegroups,
although the 3.43$\mum$ feature could also be 
due to anharmonicity and/or superhydrogenation
of the aromatic C--H stretch
(Barker et al.\ 1987, Bernstein et al.\ 1996).
Let $I_{3.4}$ and $I_{3.3}$ be
the observed intensities of the 3.43$\mum$ 
and 3.3$\mum$ emission features, respectively.
For the HD\,34700 disk, the \textit{IRTF}/SpeX
spectrum gives $I_{3.4}/I_{3.3}\approx0.49$.
Let $A_{3.4}$ and $A_{3.3}$
be the band strengths of the aliphatic 
and aromatic C--H bonds, respectively. 
Yang et al.\ (2013) calculated 
$A_{3.4}/A_{3.3}\approx1.76$.
By assuming that the 3.43$\mum$ emission is 
exclusively due to aliphatic C--H 
(i.e., neglecting anharmonicity 
and superhydrogenation), we can place 
an upper limit on the ratio of
the number of C atoms in
aliphatic sidegroups 
to that in aromatic benzene rings:
$N_{\rm C,aliph}/N_{\rm C,arom}
\approx 0.3\times\left(I_{3.4}/I_{3.3}\right)
\times\left(A_{3.3}/A_{3.4}\right)
\approx0.22$.
It is assumed here that one aliphatic C atom 
corresponds to 2.5 aliphatic C--H bonds 
and one aromatic C atom corresponds to
0.75 aromatic C--H bond (see Li \& Draine 2012).
Similarly, we can place an upper limit of 
$N_{\rm C,aliph}/N_{\rm C,arom}\approx0.14$
from the ratio of the observed intensity 
of the 6.89$\mum$ feature 
to that of the 7.7$\mum$ feature
%in the {\it Spitzer}/IRS spectrum
($I_{6.89}/I_{7.7}\approx0.064$;
see Section 3 of Li \& Draine 2012 for details).

It is not clear why PAHs are rarely seen in
debris disks. This may be related to the disk 
structure and the stellar age. 
%and the effective temperature of the star. 

There exists evidence that most of the protoplanetary disks in 
which the PAH emission features have been detected have a flaring 
disk geometry (Meeus et al.\ 2001, Acke \& van den Ancker 2004). 
A common interpretation for this is that flared disks intercept 
more stellar radiation than flat ones, especially at large distances 
from the central star; the PAH emission originates in the surface 
layers of a flared disk, where PAHs are directly exposed
to starlight (Chiang \& Goldreich 1997, Meeus et al.\ 2001,
Acke \& van den Ancker 2004, Habart, Natta, \& Kr\"ugel 2004).
It is not clear if this scenario applies to the HD\,34700 disk
as it is not spatially resolved.

If the PAHs seen in debris disks originate from 
the outgassing of cometary bodies
(i.e., the sublimation of the icy mantles 
coated on cometary grains onto which PAHs 
have condensed during the protostellar dense
cloud phase, see Li \& Lunine 2003a),
it is more likely to see PAHs in 
relatively young debris disks
as they are subject to massive
bombardment by cometary bodies
(Chyba et al.\ 1994).
With an age of a few tens of mega-years 
(Torres 2004), the HD\,34700 disk is probably still 
in the heavy bombardment phase
as it lasted more than 500 mega-years 
in our early solar system 
(Chyba et al.\ 1994).

%%%%%%%%%%%%%%% Figure 6: PAH destruction %%%%%%%%%%%%%%%
\begin{figure}[tbp]
\epsscale{0.50}
\plotone{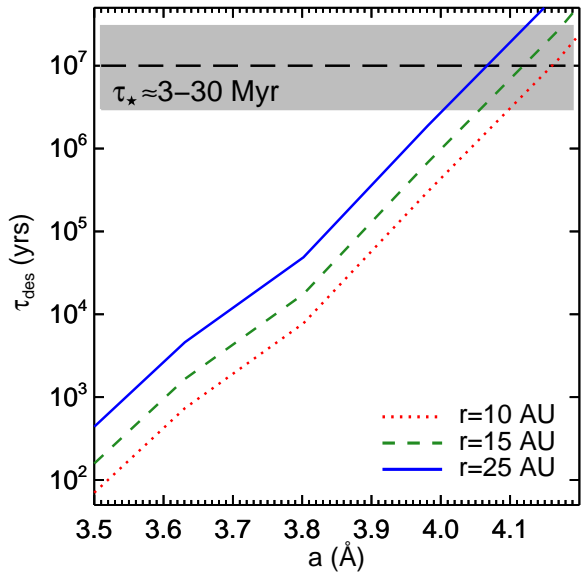}
\caption{\footnotesize
         \label{fig:kdes}
         Photodestruction timescales 
         ($\tau_{\rm des}$) for PAHs 
         at $r=10\AU$ (dotted line),
         $15\AU$ (dashed line), and
         $25\AU$ (solid line)
         as a function of size. 
         The age of HD\,34700 ($\tau_\star$)
         is rather uncertain, ranging from 
         $\sim3\Myr$ up to several tens of Myrs. 
         The age range ($\tau_\star\approx3-30\Myr$) 
         is designated by the {\bf shaded} region 
         with the representative value
         of $\tau_\star\approx10\Myr$ 
         plotted as a long-dashed line. 
         }
\end{figure}
%%%%%%%%%%%%%%% Figure 6: PAH destruction %%%%%%%%%%%%%%%

\subsection{PAH Destruction}\label{sec:dest}
A PAH molecule is subjected to photodissociation 
upon absorbing an energetic photon.
During this process, PAHs, preferentially small PAHs, 
might lose a hydrogen atom,
a hydrogen molecule, 
or an acetylene molecule (C$_2$H$_2$),
which eventually results in 
the destruction of a PAH molecule.
Following Li \& Lunine (2003a), 
we have calculated the photodestruction rates 
($k_{\rm des}$) for the PAHs in the HD\,34700 disk 
assuming that a PAH molecule is destroyed 
through the C$_2$H$_2$ ejection.
Figure~\ref{fig:kdes} shows 
the photodestruction timescales 
($\tau_{\rm des}\equiv 1/k_{\rm des}$)
at given distances ($r=10$, 15, and 25\,AU).
During the lifetime of HD\,34700 
($\tau_\star\approx10\Myr$), 
PAHs with a size of $\lesssim4.1\Angstrom$
(corresponding to PAHs with $\sim35$ carbon atoms) 
are likely to be photodestroyed.
Against the complete removal of 
such small PAHs in the HD\,34700 disk, 
these PAHs have to be replenished continuously, 
which is indeed required to account for 
the observed PAH emission. 
Assuming the PAHs in the HD\,34700 disk
follow the same spatial distribution 
as porous dust (see Equation\,(\ref{eq:dndr})) 
but with a cutoff at $r=25\AU$,
we estimate the PAH mass replenishment rate to be 
$\dot{M}_{\rm PAH}\approx4.5\times10^{-9}\Mearth\yr^{-1}$
taking only photodestruction into account
(see also Section \ref{sec:rppr}).

%%%%%%%%%%%%%%% Figure 7: RPPR %%%%%%%%%%%%%%%
\begin{figure}[tbp]
\epsscale{0.50}
\plotone{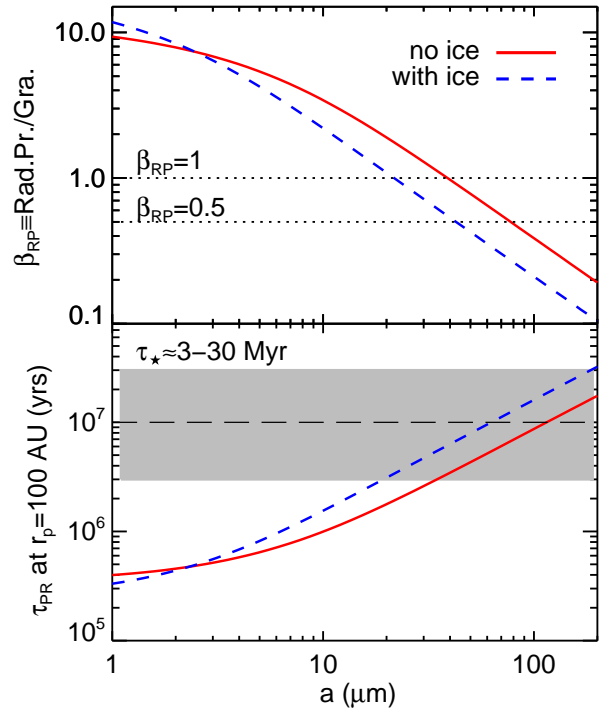}
\caption{\footnotesize
         \label{fig:rppr}
          Ratio of the radiative force 
          to the gravitational force 
          ($\beta_{\rm RP}$, top panel)
       	  and the Poynting--Robertson drag timescale 
          ($\tau_{\rm PR}$, bottom panel)
	  for ice-free porous dust ($P=0.9$: solid line) 
          and ice-coated porous dust
	  ($P^{\prime}\approx0.73$, dashed line)
          at $r=100\AU$ (note $\tau_{\rm PR}\propto r^2$). 
	  The stellar age ($\tau_\star$)
          is marked in the same way as Figure~\ref{fig:kdes}.
          }
\end{figure}
%%%%%%%%%%%%%%% Figure 7: RPPR %%%%%%%%%%%%%%%

\subsection{Radiation Pressure}\label{sec:rp} 
Small dust in a debris disk is continuously 
expelled outward due to the RP 
against the gravitational force 
(Backman \& Paresce 1993).
Following the approach described in Section 4.3 
of Li \& Lunine (2003a),
we derive the ratio of the radiative force 
to the gravitational force ($\beta_{\rm RP}$)
as a function of grain size.%
Both the RP and the gravitational force are
inversely proportional to $r^2$, 
as a result, $\beta_{\rm RP}$ is
independent of the distance from the central star.

We find that grains smaller than 
$\simali$20--40$\mum$ have $\beta_{\rm RP}\ga1$
(see Figure~\ref{fig:rppr}).
This indicates that grains with a radius 
smaller than $\simali$40$\mum$
($\simali$20$\mum$ for icy grains) 
will be blown out due to RP.
Moreover, assuming that the gas drag 
is not overwhelming which is true for
debris disks, 
even larger grains ($\simali$40--80$\mum$)
could be pushed outward 
(see Figure~\ref{fig:rppr})
because dust particles
released from parent bodies at periastron
only need $\beta_{\rm RP}=0.5$ for ejection
(Burns et al.\ 1979).

Unlike dust grains, the $\beta_{\rm RP}$ ratio 
for PAHs is not sensitive to their sizes 
as they are in the Rayleigh limit (see Li 2009a).
We estimate $\beta_{\rm RP}\approx35$
for the PAHs in the HD\,34700 disk. 
%and $\tau_{\rm PR}\approx6000\yr$ for PAHs at $r=25\AU$.

\subsection{Poynting--Roberson Drag}\label{sec:pr}
%
 
%Besides the radiation pressure, 
Dust grains can migrate inward 
due to the Poynting--Roberson (PR) drag
(Backman \& Paresce 1993).
We calculate the PR drag timescale ($\tau_{\rm PR}$), 
which is a function of grain size 
and the distance ($r$) from the star.
Note that, unlike the RP ejection,
the PR drag is proportional to $r^2$.

%For instance, 
At the peak distance of $\rp\approx100\AU$, 
grains smaller than $\simali$50--100$\mum$
have a PR lifetime shorter than 
the age of HD\,34700 
(assuming $\tau_\star\approx10\Myr$;
see Figure~\ref{fig:rppr}).
The closer dust grains are located to the star,
the bigger grains will be gradually accreted to 
the central star.
For PAHs, we estimate 
$\tau_{\rm PR}\approx6000\yr$ for PAHs at $r=25\AU$,
which is independent of their sizes 
for the same reason as for $\beta_{\rm RP}$.

\subsection{Dominant Processes for the Removal of 
            Dust and PAHs in the HD\,34700 Disk}
            \label{sec:rppr}
Taking into account 
the RP ejection, 
the PR drag, 
as well as the photodissociation of PAHs,
we examine the removal of dust and PAHs 
in the HD\,34700 disk
as a function of their sizes 
and distance from the star.
The RP timescale in the HD\,34700 disk
is estimated to be $\tau_{\rm RP}(r)
\approx912\left(r/100\,{\rm AU}\right)^{3/2}\yr$
assuming that it is comparable to the local
dynamical timescale $\tau_{\rm dyn}=2\pi/\Omega(r)$,
where $\Omega(r)$ is the Keplerian frequency.
For grains with $\beta_{\rm RP}\gtsim0.5$
(i.e., $a\la80\mum$),
the RP timescale at a given distance 
is much smaller than the PR drag timescale
(see Figure \ref{fig:rppr}). 
Therefore, for these grains 
%with $\beta_{\rm RP}\ga1$ (i.e., $a\la40\mum$)
the RP expulsion is the dominant removal process 
over the entire disk.
For larger grains (i.e., $a\ga80\mum$),
if the PR timescale $\tau_{\rm PR}(a, r)$ is 
shorter than the stellar age $\tau_\star$,
the PR drag becomes an effective removal mechanism.
At $r=100\AU$,
this applies to grains of $a\la100\mum$ 
(see Figure~\ref{fig:rppr}).
For grains at a larger distance from the star,
the size range over which the PR drag is effective
becomes narrower, and eventually, the PR drag 
no longer contributes to the dust removal 
beyond $r\sim150\AU$.  

Since the overall dust removal in the HD\,34700 disk 
is dominated by the RP ejection,
we calculate the RP dust mass-loss rate 
for our best-fit model.
Following Equation (20) in Li \& Lunine (2003a), 
we derive
$\dot{M}_{\rm RP}\approx2.05\times10^{-4}\Mearth\yr^{-1}$.
Hence, $\simali$2000$\Mearth$ of dust 
has been removed during the lifetime of HD\,34700. 
For comparison,
the PR dust mass-loss rate 
integrated over the dust size distribution
and over the entire disk 
is estimated to be 
$\dot{M}_{\rm PR}\approx1.78
\times10^{-7}\Mearth\yr^{-1}$. 

In the same manner as for dust grains, 
we estimate the PAH mass loss rate 
due to the RP expulsion 
to be $\dot{M}_{\rm RP}({\rm PAH})
\approx2.92\times10^{-6}\Mearth\yr^{-1}$.
Note that PAHs are also subject to 
photodissociation
(see Section \ref{sec:dest}).
We found that the RP timescale is much shorter
than the photodestruction timescale
except for the smallest PAHs 
($a$\,$\approx$\,3.5--3.6$\Angstrom$).
The PAH mass loss rate 
via the RP expulsion ($\dot{M}_{\rm RP}$)
yields that $\simali$30$\Mearth$ 
of PAH molecules have been removed 
from the disk or have been replenished 
to maintain the current PAH abundance during 
the 10\,Myr lifetime of HD\,34700.

It is interesting to note that 
the PAH-to-dust mass-loss ratio,
$\dot{M}_{\rm RP}({\rm PAH})/\dot{M}_{\rm RP}({\rm dust})
\approx1.4\times10^{-2}$,
is about three order of magnitudes
higher than the PAH-to-dust mass ratio
of $M_{\rm PAH}/M_{\rm dust}
\approx1.3\times10^{-5}$ 
(see Section~\ref{sec:pah}).
This is because only grains with $a\la80\mum$
will be ejected by the RP expulsion 
(see Figure \ref{fig:rppr})
and with a best-fit dust size distribution
of $dn/da\propto a^{-3.5}$
(see Section~\ref{sec:results}), 
the total dust mass 
in the HD\,34700 disk
is dominated by much larger grains.
Thus, only a small mass fraction of 
the dust is subject to the RP expulsion, 
whereas PAHs of all sizes 
have been removed or replenished. 

%
%In spit of the RP ejection, however,
%the disk can contain a stable 
%amount of small grains through
%replenishment by collisions of larger bodies 
%(Krivov et al.\ 2000).

The high removal rates of small dust and PAHs 
derived above require an efficient replenishment 
probably provided by the outgassing and collisions 
of planetesimals, asteroidal, and cometary bodies 
to contain a stable amount of small grains and PAHs
in the disk (Backman \& Paresce 1993, Wyatt 2008, 
Kiefer et al.\ 2014).
In particular, as PAHs have been detected in 
interplanetary dust particles,
comets, and meteorites 
(e.g., see Li 2009b for overview),
cometary origin of PAHs 
in the HD\,34700 disk is plausible. 
By systematically studying 
all debris disks 
with detected PAH features, 
we will (hopefully) be able to constrain 
the origin of PAHs and dust in debris disks
(J. Y.~Seok \& A.~Li, 2015, in preparation).

\section{Summary}\label{sec:summary}
The HD\,34700 disk is one of 
the few debris disks that show 
prominent PAH emission in their IR spectra.
We have modeled the PAH emission of 
the HD\,34700 disk as well as its entire SED 
from the near-IR up to submm/mm wavelength regime.
The stellar RP 
and Poynting--Robertson drag
quickly remove the dust and PAHs
in the HD\,34700 disk, implying that
they must have been replenished 
continuously
to maintain their current abundances.

The major PAH features at 3.3, 6.2, 7.7, 8.6, and 11.3$\mum$
(as well as the minor features at 3.43, 6.89, and 7.23$\mum$)
are present in the IR spectra, indicating the presence of
a mixture of neutral and ionized PAHs 
(with aliphatic sidegroups) in the disk.
By comparing the photoionization and electron
recombination timescales, it is found that 
the PAHs in the inner disk are likely to be ionized 
while those in the outer disk are to be neutral
or negatively charged. 
Also, at a given distance from the star,
smaller PAHs are more likely to be neutral
whereas larger PAHs are more likely to be charged.
Hence, one would expect a mixture of
neutral and ionized PAHs in the HD\,34700 disk.

Adopting a log-normal size distribution 
for the PAHs, the HD\,34700 disk is
in favor of smaller PAHs
($a_{0}=3.5\Angstrom$ and $\sigma=0.3$)
compared to that of the diffuse ISM.
This is supported by the the prominent 
3.3$\mum$ PAH feature,
which is indicative of 
small, neutral PAHs. 
%in the HD\,34700 disk.
%
The 3.43$\mum$ feature commonly attributed
to aliphatic C--H stretch allows us
to place an upper limit of $\simali$0.22 
on the ratio of C atoms in aliphatic sidegroups
to that in aromatic benzene rings. 
The 6.89$\mum$ feature, 
an aliphatic C--H deformation band,
derives the ratio of aliphatic C atoms
to aromatic C atoms to be $\simali$0.14.

The total IR luminosity of 
$L_{\rm IR}\approx2.75 \Lsun$
is derived with a distance of $d=260$ pc.
This leads to $L_{\rm IR}/\Lstar\approx0.13$, 
and the PAH emission accounts for
$\simali$10\% of the total $L_{\rm IR}$.
Our best fit model suggests
that the dust spatial distribution peaks at
$\approx100\AU$ and decreases outward
with a power-law index of $\gamma\approx2$.
Taking the disk geometry into account,
the total dust mass in the disk is estimated
to be $M_{\rm dust}\approx21.9\Mearth$.
Porous grains larger than
$\simali$30$\mum$ attain an equilibrium 
temperature of $T\ltsim$120$\K$ 
at $r\approx25\AU$,
indicating that ice mantles on these grains
begin to develop at $r\approx25\AU$.
Therefore,
$r=25\AU$ is regarded as the ``snowline''
in the HD\,34700 disk.

Without any prior knowledge of 
the PAH spatial distribution,
the observed PAH emission yields 
a total mass of 
$M_{\rm PAH}\approx4.53\times10^{-7}
\left(r/{\rm AU}\right)^{2}\Mearth$
if one assumes that all the PAHs are
at a distance of $r$ from the central star.
As free-flying PAHs will easily condense onto
the ice mantles beyond the snowline,
an upper limit of 
$M_{\rm PAH}\approx2.83\times10^{-4}\Mearth$
is derived for $r=25\AU$.
This results in a PAH-to-dust mass ratio of
$M_{\rm PAH}/M_{\rm dust}\approx1.3\times10^{-5}$,
indicating a deficiency of PAHs in debris disks
compared with the diffuse ISM.

The photodissociation of PAH molecules 
occurs continuously 
by absorbing an energetic photon,
and small PAHs ($\lesssim4.1\Angstrom$) 
are likely to have been photo-destroyed 
during the lifetime of HD\,34700
($\tau_\star\approx10\Myr$).
We calculate the PAH mass replenishment rate 
to be $\approx4.5\times10^{-9}\Mearth\yr^{-1}$ 
due to photodissociation.

The stellar RP dominates 
the dust and PAH removal in the HD\,34700 disk.
Due to the RP
together with Poynting--Robertson drag,
$\simali$2000$\Mearth$ of dust has been removed 
during the $\simali$10$\Myr$ lifetime of HD\,34700.
Similarly, $\simali$30$\Mearth$ of PAHs has been 
removed out of the disk.
Collisions and/or sublimations of planetesimals, 
asteroids, and comets are the likely origin of 
the dust and PAH replenishment.

%Gaia will provide more accurate distance to HD 34700, and
%JWST will provide more IR data.

\acknowledgements
We thank L. D.~Keller, S.~Wang
and the anonymous referee
for their very helpful suggestions.
We are supported in part by
NSF AST-1109039, 
NNX13AE63G, 
%NSFC\,11173007, 
NSFC\,11173019, 
%NSFC\,11273022,
and the University of Missouri Research Board.

%%%%%%%%%%%%%%% References %%%%%%%%%%%%%%%%%%%%%%%%%%%%

\end{document}